\documentclass[]{scrartcl}

\usepackage[utf8]{inputenc}
\usepackage{graphicx}
\usepackage{amsmath}
\usepackage{amsfonts}
\usepackage{xcolor}
\usepackage[normalem]{ulem}
\usepackage{enumitem}
\usepackage{booktabs}
\usepackage[super,compress]{natbib}
\usepackage{parskip}
\usepackage{acronym}
\usepackage{rotating}
\usepackage{comment}
\usepackage{multirow}
\usepackage{array}
\usepackage{authblk}
\usepackage{url}

\newcommand{\bs}[1]{\boldsymbol{{#1}}}
\newcommand{\ea}[1]{\emph{\textbf{#1}}}
\newcolumntype{P}[1]{>{\raggedleft\arraybackslash}p{#1}}

\title{Implications for HIV elimination by 2030 of recent trends in undiagnosed infection in England: an evidence synthesis}

\author[1,*]{Anne M Presanis}
\author[2]{Peter Kirwan}
\author[3,2]{Ada Miltz}
\author[2]{Sara Croxford}
\author[2]{Ross Harris}
\author[2]{Ellen Heinsbroek}
\author[1]{Chris Jackson}
\author[2]{Hamish Mohammed}
\author[2]{Alison Brown}
\author[2]{Valerie Delpech}
\author[2]{O Noel Gill}
\author[1,2]{Daniela De Angelis}
\affil[1]{Medical Research Council Biostatistics Unit, University of Cambridge}
\affil[2]{Public Health England}
\affil[3]{Institute of Global Health, University College London}
\affil[*]{\small Corresponding author}

\date{\today}

\begin{document}

\maketitle

\newpage

\section*{Abstract}

\paragraph{Background:} A target to eliminate Human Immuno-deficiency Virus (HIV) transmission in England by 2030 was set in early 2019. Estimates of recent trends in HIV prevalence, particularly the number of people living with undiagnosed HIV, by exposure group, ethnicity, gender, age group and region, are essential to monitor progress towards elimination.
\vspace{-5mm}
\paragraph{Methods:} A Bayesian synthesis of evidence from multiple surveillance, demographic and survey datasets relevant to HIV in England is employed to estimate trends in: the number of people living with HIV (PLWH); the proportion of these people unaware of their HIV infection; and the corresponding prevalence of undiagnosed HIV. All estimates are stratified by exposure group, ethnicity, gender, age group (15-34, 35-44, 45-59, 60-74), region (London, outside London) and year (2012-2017).
\vspace{-5mm}
\paragraph{Findings:} The total number of PLWH aged 15-74 in England increased from 82,400 (95\% credible interval, CrI, 78,700 to 89,100) in 2012 to 89,500 (95\% CrI 87,400 to 93,300) in 2017. The proportion diagnosed steadily increased from 84\% (95\% CrI 77 to 88\%) to 92\% (95\% CrI 89 to 94\%) over the same time period, corresponding to a halving in the number of undiagnosed infections from 13,500 (95\% CrI 9,800 to 20,200) to 6,900 (95\% CrI 4,900 to 10,700). This decrease is equivalent to a halving in prevalence of undiagnosed infection and is reflected in all sub-groups of gay, bisexual and other men who have sex with men and most sub-groups of black African heterosexuals. However, decreases were not detected for some sub-groups of other ethnicity heterosexuals, particularly outside London.
\vspace{-5mm}
\paragraph{Interpretation:} In 2016, the Joint United Nations Programme on HIV/ AIDS target of diagnosing 90\% of people living with HIV was reached in England. To achieve HIV elimination by 2030, current testing efforts should be enhanced to address the numbers of heterosexuals living with undiagnosed HIV, especially outside London.

\section{Introduction}

Early diagnosis of \ac{HIV} infection and access to \ac{ART} can 
prevent onward transmission \cite{Cohen2011}. Regular assessment of the burden of \ac{HIV} 
is essential for 
evaluation of public health policies aimed at reducing transmission, such as 
``\ac{TasP}'' \cite{GranichEtAl2009,UNAIDS9090902014,Churchill2016} and \ac{PrEP} \cite{McCormack2016}. 
Knowledge of the number and proportion of \ac{HIV} infections remaining undiagnosed is crucial for monitoring progress towards elimination of \ac{HIV} transmission by 2030 \cite{UNAIDS9090902014}.
Since undiagnosed cases are inherently unobservable, these quantities must be estimated.

Annual \ac{UK} estimates of \ac{HIV} prevalence, both diagnosed and undiagnosed, the number of \ac{PLWH} and the proportions of infections that are diagnosed are published 
\cite{PHEHIVreport2018} for \ac{MSM}, \ac{PWID}, and heterosexual individuals of both black African  and other ethnicity (as self-reported to the \ac{UK} Census of 2011 \cite{ONSCensus2012}). 
The estimates 
are the yardstick for measuring the \ac{UK}'s progress towards the first \ac{UNAIDS} 90-90-90 target, i.e. 90\% of \ac{HIV} infections diagnosed, and for informing \ac{HIV} testing guidelines \cite{BHIVA2008,NICE2017} and prevention campagins \cite{NAT2013,HPEcampaigns2019}. Since 2005, these estimates of \ac{HIV} prevalence, seen as cross-sectional ``snapshots'' of the state of the epidemic, have been derived through a ``\ac{MPES}'', a statistical approach combining and triangulating multiple sources of surveillance and survey data \cite{Goubar2008,DeAngelis2014}. Data on exposure group sizes, numbers diagnosed and in care, and \ac{HIV} prevalence from prevalence surveys and 
testing data, are synthesised to estimate the undiagnosed fraction. As both the epidemic and the available data sources have changed over time \cite{PHEHIVreport2018}, the \ac{MPES} model has also evolved structurally since its creation, to make greater and more efficient use of the available data.

This paper presents a major extension to our MPES model that integrates sequential cross-sectional estimates to produce trends (with credible intervals) from 2012-2017 in \ac{HIV} prevalence and the number undiagnosed, by route of probable exposure (exposure group), ethnicity, gender, age and region.

\section{Methods \label{sec:Methods}}

The adult population of England in the years 2012 to 2017 was stratified by exposure group and ethnicity (\ac{MSM}; \ac{PWID}; black African heterosexuals; other ethnicity heterosexuals); gender (men, women); age (15-34, 35-44, 45-59, 60-74); and region (London, England outside London). The \ac{MSM} and heterosexual exposure groups are further sub-divided by whether or not they had attended a sexual health clinic in the last year for a \ac{STI}-related need (``recent clinic attendee''). \ac{PWID} are stratified by whether or not they had injected drugs in the last year.

\subsection{Multi-parameter evidence synthesis}

To estimate 
\ac{HIV} prevalence in each stratum, data 
are combined with prior assumptions, in a Bayesian model that encodes the relationships between the data from each source and the quantities to be estimated. The \ac{MPES} approach \cite{Goubar2008,DeAngelis2014} consists of: defining the key quantities (`basic parameters') to be estimated, with any prior knowledge of these quantities summarised in a \emph{prior distribution}; relating mathematically the quantities (`functional parameters') directly informed by each data source to the basic parameters, defining the likelihood of the data
; updating the prior distribution with our current knowledge, summarised by the likelihood, to obtain a \emph{posterior distribution} of 
all basic and 
functional parameters, that summarises all uncertainty in both 
data and 
parameters. Any 
unobserved functional parameters of interest
are also derived as functions of the basic parameters. This method ensures that resulting estimates are consistent with all included data and model assumptions. 
We use Markov chain Monte Carlo
to 
draw samples from the posterior distribution, 
summarised by their median and a $95\%$ credible interval (\ac{CrI}) defined by the $2.5$ and $97.5$ percentiles. Posterior probabilities of either an increase or decrease over 2012-2017 in 
each outcome are also calculated, as the proportion of posterior samples that are greater/smaller in 2017 compared to 2012. 
All analyses were carried out in R version 3.4.4, 
\texttt{rstan} and Stan \cite{rStan2018}.

\subsection{Model}

We 
estimate three basic parameters for each stratum $asrt$ defined by age group $a$, gender $s$, region $r$ and year $t$:
\begin{description}
    \item[$\rho_{agsrt}$] the proportion of the population in stratum $asrt$ who are in exposure group $g$;
    \item[$\pi_{agsrt}$] 
    \ac{HIV} prevalence in stratum $agsrt$;
    \item[$\delta_{agsrt}$] the proportion of \ac{HIV} infections in stratum $agsrt$ that are diagnosed.
\end{description}

Given knowledge 
of these three basic parameters per stratum, 
any 
functional parameter
can be derived by defining it in terms of the basic parameters. The key functional parameters 
are: the number of \ac{PLWH} in each stratum, $N_{asrt}\rho_{agsrt}\pi_{agsrt}$, where $N_{asrt}$ is the total population in stratum $asrt$, 
obtained from \ac{ONS} data; the 
number of undiagnosed infections, $N_{asrt}\rho_{agsrt}\pi_{agsrt}\left( 1 - \delta_{agsrt} \right)$; and the corresponding prevalence of undiagnosed infection, $u_{agsrt} = \pi_{agsrt}\left( 1 - \delta_{agsrt} \right)$.

\subsection{Data and model assumptions}

A substantial range of evidence is available in England to inform both exposure group sizes and HIV prevalence, either directly or indirectly. 
Here we summarise the data and their 
 relationships to the parameters, while full details and a schematic diagram of the model are given in the Appendix. 

\paragraph{Group sizes}
Estimates of the size of the total population of England, by age, gender and region strata
, are available from the \ac{ONS}, for each year \cite{ONS_MYEs2018}. 
The 2011 Census 
\cite{ONSCensus2012} 
provides information on 
the proportions of the population self-reporting their ethnicity (black African or any other ethnicity). These proportions are applied to the heterosexual (non-\ac{MSM}, non-\ac{PWID}) group each year, to derive the annual distributions by ethnicity of the heterosexual groups.

Survey-weighted estimates of the proportion of men who are \ac{MSM}, by age and region, are available from the \ac{NATSAL}, a national stratified probability sample 
from 2011 \cite{Mercer2016}.

Information on \ac{PWID} population sizes, both current and ex, is available from previous studies, based on data from 2005-2012 \cite{HayEtAl2011,KingEtAl2014,SweetingEtAl2009}. Group sizes are therefore estimated 
for 2012, and are assumed not to change over time. 

The sizes of the sub-groups of \ac{MSM} and heterosexuals by ethnicity who have attended a \ac{SHS} for a \ac{STI}-related need in the last year, for each year 2012-2017, are directly available from the GUMCAD \ac{STI} surveillance system, a disaggregated, pseudonymised data return submitted by all commissioned \ac{SHS}s across England \cite{Savage2014}.

Table 2 of the Appendix 
gives details of the group size parameters, their prior distributions or functional forms, and which datasets inform them.

\paragraph{Prevalence in clinic-attending groups}
The \ac{SHS} data provide indirect evidence on \ac{HIV} prevalence, both diagnosed and undiagnosed, in the clinic-attending groups for each year 2012-2017. Four components of \ac{HIV} prevalence can be derived from the clinic data combined with model assumptions: previously diagnosed prevalence ($g_{agsrt1}$); newly diagnosed prevalence ($g_{agsrt2}$); prevalence of undiagnosed infection in those not 
offered a \ac{HIV} test ($g_{agsrt3}$); and prevalence of undiagnosed infection in those opting out of a \ac{HIV} test ($g_{agsrt4}$) (Table 
3 of the Appendix). The diagnosed components are directly observed, whereas the undiagnosed prevalences are estimated by relating them to newly diagnosed prevalence. 

\paragraph{Undiagnosed prevalence in \ac{MSM}}
The \ac{GMSHS} \cite{Aghaizu2016}
samples \ac{MSM} at community venues 
in London every 2-3 years. 
Participants are offered a \ac{HIV} test and asked about recent sexual health service attendance and any previous \ac{HIV} diagnosis. Since \ac{GMSHS} participants may be higher risk than average \ac{MSM}, this source provides over-estimates of \ac{HIV} prevalence in 
all \ac{MSM}. 
The data are therefore used indirectly, to inform the odds ratio of (previously) undiagnosed prevalence in recent clinic-attending vs non-clinic-attending\ac{MSM} 
(Table 
3 of the Appendix).

\paragraph{Prevalence and proportion diagnosed in \ac{PWID}}
The \ac{UAM} \cite{PublicHealthEngland2018} annual survey of \ac{PWID} recruits attendees at needle exchanges, methadone treatment and other drug services and involves a (self-reported) questionnaire and dried blood spot, which is tested for HIV antibodies. The survey provides information on both \ac{HIV} prevalence, $\pi_{agsrt}$, and the 
proportion of \ac{PWID} who have ever had a \ac{HIV} diagnosis, $\delta_{agsrt}$, for each year 2012-2017. The sampled population is assumed to represent current \ac{PWID}. 

For the ex-\ac{PWID} group, we assume their stratum-specific proportion diagnosed is larger than for 
current \ac{PWID} 
(Table 
3 of the Appendix). 

\paragraph{Previously undiagnosed prevalence in women}
The \ac{NSHPC} \cite{Peters2018} is an annual register of all \ac{HIV} diagnoses among pregnant women. 
The observed number of diagnoses occurring during current pregnancies are combined with data from \ac{ONS} on the annual number of live births \cite{ONSliveBirthsRoB2018}, to indirectly inform previously undiagnosed prevalence in non-\ac{PWID} women under 45 years of age (Table 
4 of the Appendix), stratified by ethnicity, age and region. 

\paragraph{Proportion diagnosed in black African heterosexuals}
The \ac{AHSS} \cite{Bourne2014}, carried out across England in 2014, provides information on the proportion of African \ac{PLWH} who self-report ever having had a \ac{HIV} test, $\theta_{asrt}$. As this proportion does not directly inform proportions aware of their infection status, 
we instead use the data indirectly, 
to inform the male-to-female odds ratio of the proportion diagnosed
(Table 
4 of the Appendix).

\paragraph{Prevalence in the lowest risk group}
The \ac{NHSBT} and \ac{PHE} carry out blood-borne virus testing among blood donors, a population considered at very low risk of \ac{HIV} infection, by gender, region and age \cite{PHE_NHSBT_2017}. 
The \ac{HIV} prevalence in blood donors, $\pi^{(\textsc{BD})}_{asrt}$, from the \ac{NHSBT} data are therefore used indirectly, to inform the male-to-female odds ratio of prevalence in heterosexuals who 
are not recent clinic attendees
(Table 
4 of the Appendix).

\paragraph{Numbers diagnosed}
The \ac{HARS} \cite{Rice2017} is a comprehensive surveillance system recording all new \ac{HIV} diagnoses, as well as regular follow-up reports on the clinical status of all diagnosed \ac{HIV}-positive patients attending \ac{HIV} outpatient services for clinical care. Due to the very high retention in care of \ac{HIV} patients in the UK, the \ac{HARS} data provide a complete yearly snapshot of the number of people living with diagnosed \ac{HIV} in the UK, by ethnicity, age, gender, region and year; as well as the probable \ac{HIV} exposure group distribution of these individuals 
(Table 
5 of the Appendix). 

\paragraph{Borrowing strength across strata}
As the spread of available data across different strata is uneven, we ``borrow strength'' from exposure-ethnicity-gender-age-region-year strata with more data to smooth and increase precision in estimates of both \ac{HIV} prevalence $\pi_{agsrt}$ and the proportion diagnosed $\delta_{agsrt}$ for strata with fewer data sources. This is achieved via a hierarchical random effects model that assumes that the log-odds ratios of prevalence and proportion diagnosed in non-clinic-attending groups versus clinic-attending groups might plausibly be thought similar, but not exactly equal, across strata. 
Smoothing of trends in the log-odds ratios across years is achieved by also linking years hierarchically.

\section{Results \label{sec:Results}}

\begin{figure}
\includegraphics[width = \textwidth]{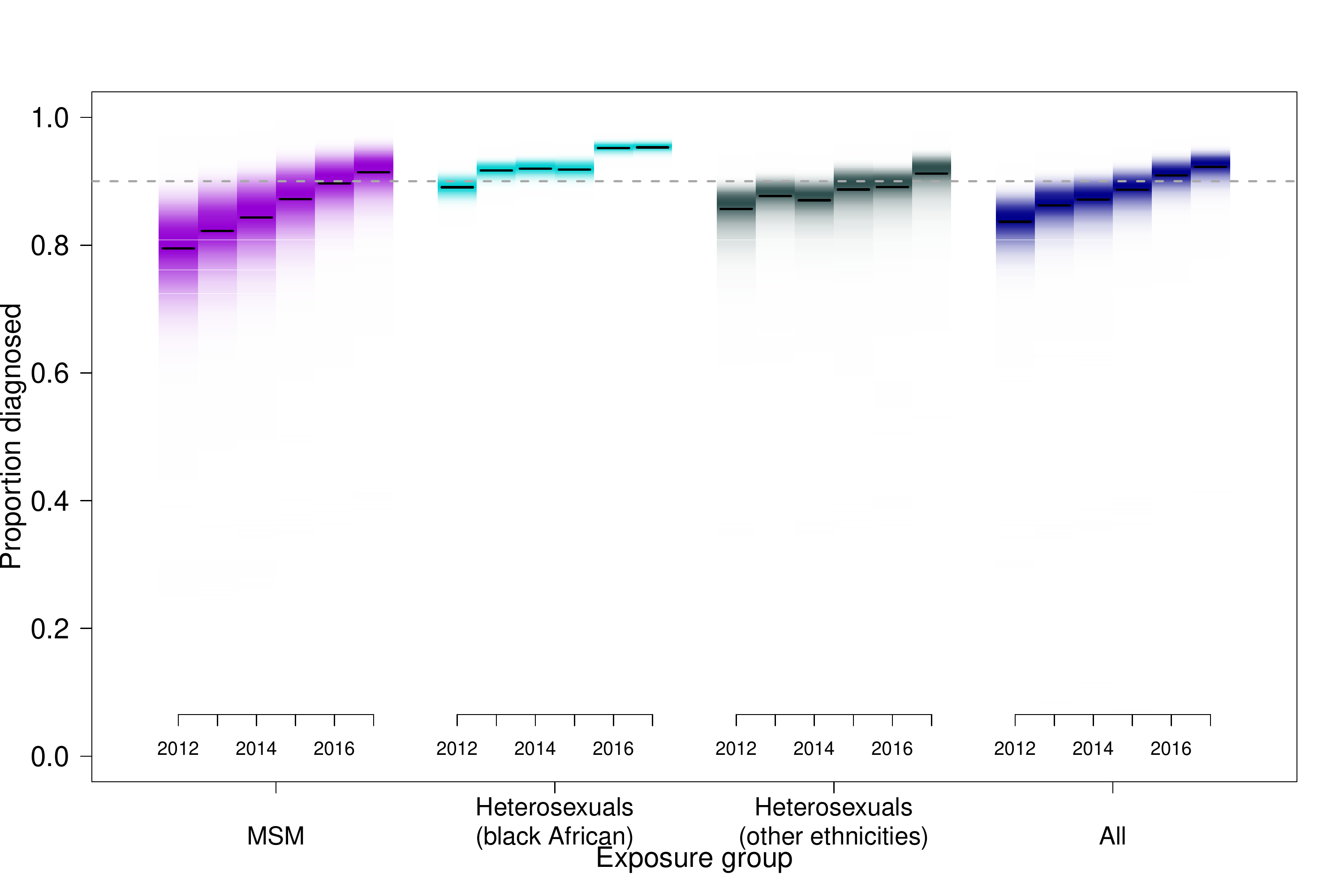}
\caption{Proportion of \ac{PLWH} whose infection is diagnosed, by exposure group and year (2012-2017). Each vertical strip is a ``density strip plot'' representing the posterior uncertainty in the estimate. The intensity of colour is proportional to the posterior density of the proportion diagnosed, i.e. darker areas are more likely than lighter. The horizontal solid black lines represent the posterior median estimate. The UNAIDS 90\% diagnosed target is shown by the horizontal dashed grey line.}
\label{fig:prDiagOverall}
\end{figure}

The estimated number of \ac{PLWH} in England aged 15-74 who were unaware of their infection decreased from 13,500 (95\% \ac{CrI} 9,800 to 20,200) in 2012 to 6,900 (95\% \ac{CrI} 4,900 to 10,700) in 2017, posterior probability of a decrease $p = 0.992$, Table \ref{tab:postEst}). This decrease corresponded to a halving in the prevalence of undiagnosed infection from 0.34 (95\% \ac{CrI} 0.25 to 0.51) to 0.17 (95\% \ac{CrI} 0.12 to 0.26) per 1,000 population over the five-year period. The decreases in numbers undiagnosed and undiagnosed prevalence are notably greater in London than outside London (Table \ref{tab:postEst}), demonstrated by the consistently lower posterior probabilities $p$ of a decrease.
An increase in the number of people living with diagnosed \ac{HIV} resulted in the total number of \ac{PLWH} increasing from 82,400 (95\% \ac{CrI} 78,700 to 89,100) to 89,500 (95\% \ac{CrI} 87,400 to 93,300, $p = 0.978$) in 2012-2017. The percentage diagnosed in the overall population therefore steadily increased from 84\% (95\% \ac{CrI} 77 to 88\%) in 2012 to 92\% (95\% \ac{CrI} 89 to 94\%) in 2017, reaching the \ac{UNAIDS} 90\% diagnosed target in 2016, and even earlier in 2013 for black African heterosexuals (Figure \ref{fig:prDiagOverall}). 

\begin{sidewaysfigure}
\includegraphics[width = \textwidth]{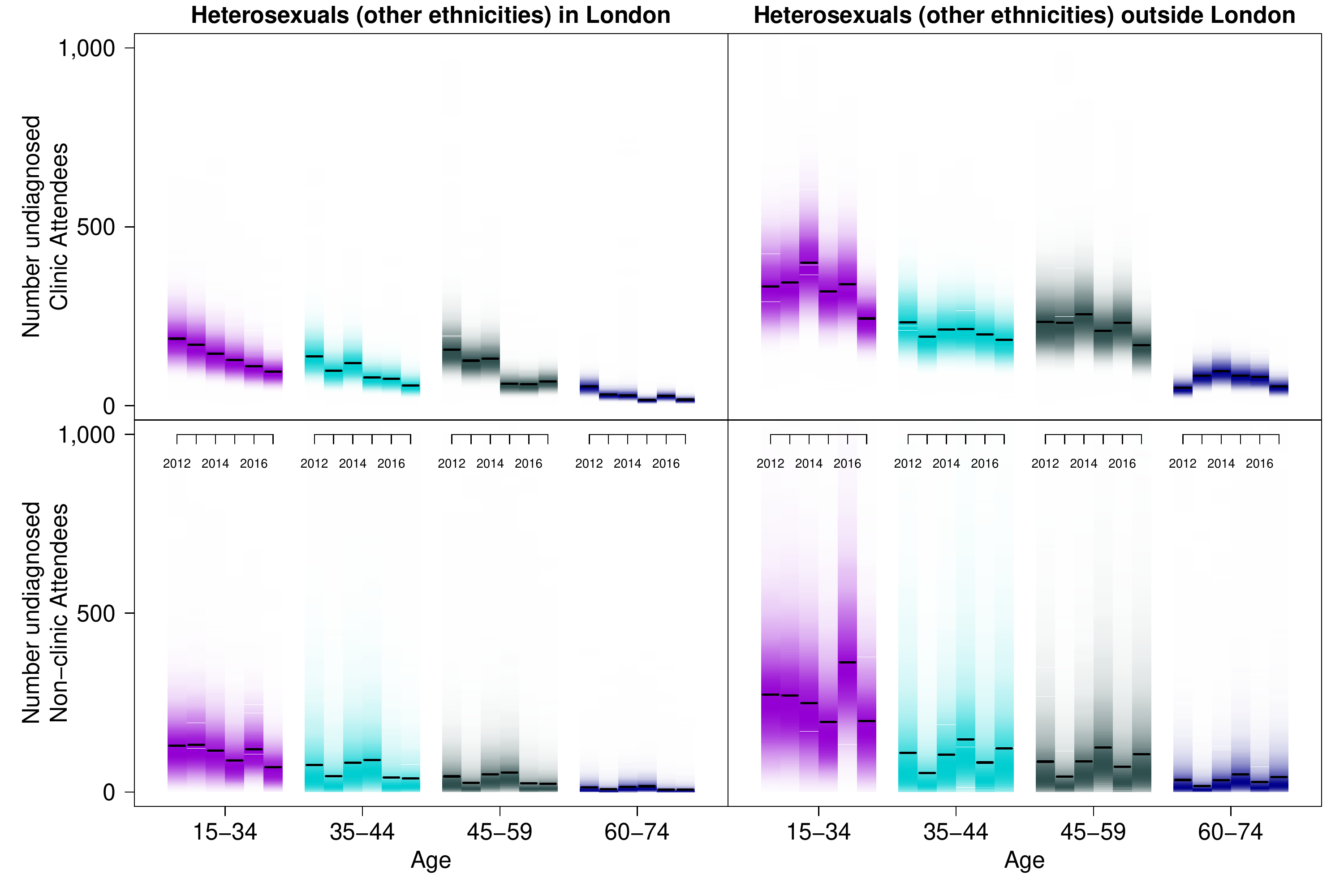}
\caption{Number of other ethnicity heterosexuals living with undiagnosed \ac{HIV}, by clinic attendance, region, age and year (2012-2017).}
\label{fig:NundiagNBAhets}
\end{sidewaysfigure}

A closer look at each sub-group reveals considerable variability in the pace of reduction of undiagnosed infections (Table \ref{tab:postEst} and Figure 5 of the Appendix). While the 90\% target was achieved for most sub-groups by 2017, including both black African and other ethnicity heterosexuals overall, the target has yet to be reached for other ethnicity heterosexuals outside London who 
were recent clinic attendees. Moreover, there is less evidence of an increase in the percentage diagnosed for this sub-group, as reflected by the uncertainty in the estimates of the number unaware of their \ac{HIV} infection (Figure \ref{fig:NundiagNBAhets}). In total the numbers of other ethnicity heterosexuals living with undiagnosed \ac{HIV} reduced from 2,300 (95\% \ac{CrI} 1,700 to 4,000) to 1,600 (95\% \ac{CrI} 1,100 to 3,200, $p = 0.857$) over 2012-2017 -- with a corresponding drop in prevalence of undiagnosed infection from 0.059 (95\% \ac{CrI} 0.044 to 0.104) to 0.040 (95\% \ac{CrI} 0.027 to 0.082) per 1,000 population (Table \ref{tab:postEst}). However, such a decrease was not discernible in the two older age groups outside London, particularly in those who were not clinic attendees (Table \ref{tab:postEstNBAbyAge}). Indeed, in general, posterior probabilities of a decrease were lower for those who had not recently attended a clinic than for those who had. 
Despite evidence for decreases in the numbers of undiagnosed \ac{PLWH} among clinic-attending other ethnicity heterosexuals, 
in terms of undiagnosed prevalence (Table \ref{tab:postEstNBAbyAge}), the rates remained much larger for clinic-attendees than for 
non-attendees.

\begin{sidewaysfigure}
\includegraphics[width = \textwidth]{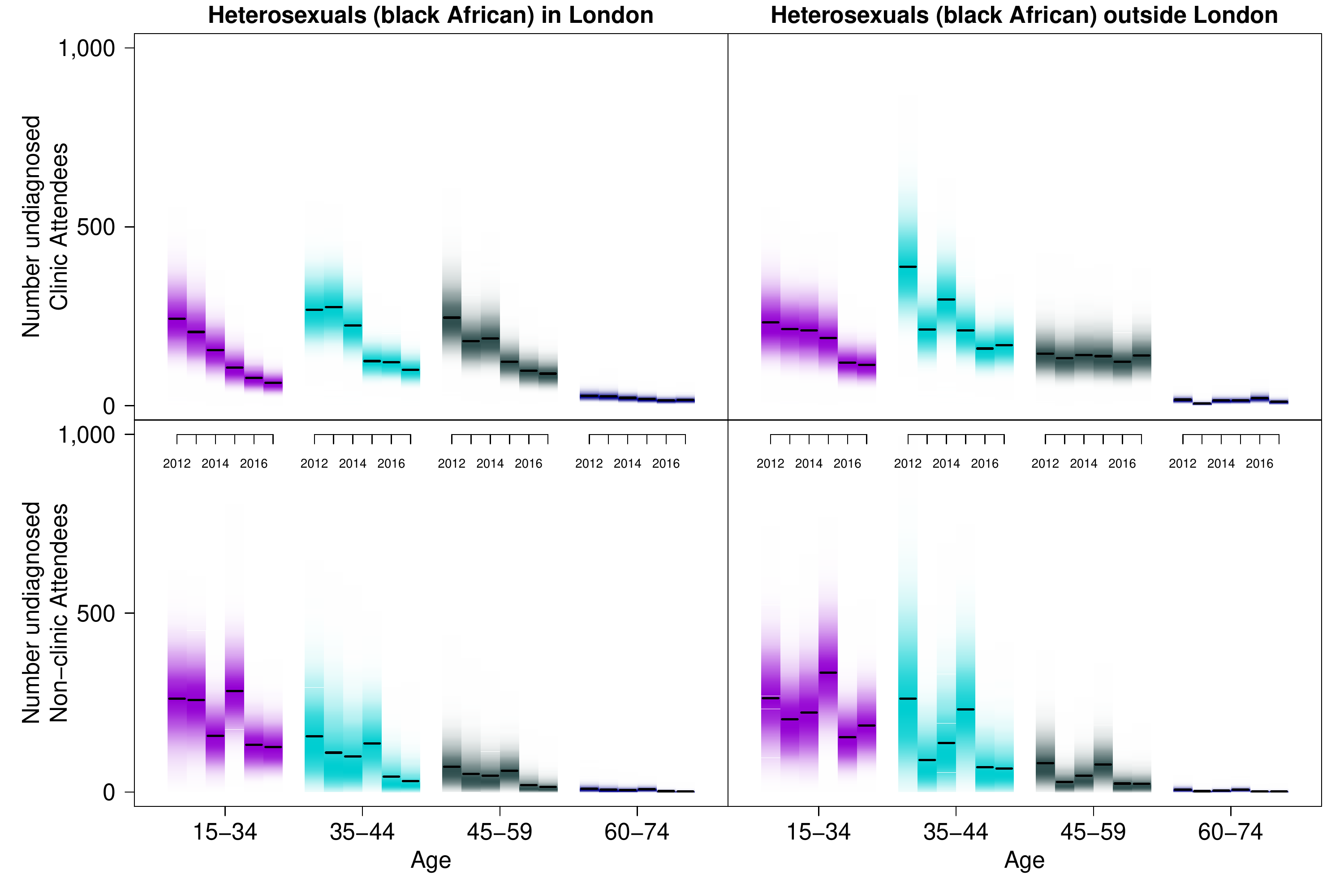}
\caption{Number of black African heterosexuals living with undiagnosed \ac{HIV}, by clinic attendance, region, age and year (2012-2017).}
\label{fig:NundiagBAhets}
\end{sidewaysfigure}

Progress in reducing the undiagnosed pool of infection among black African heterosexuals was more evident than for other heterosexuals (Figure \ref{fig:NundiagBAhets}), particularly in London. In total, the decrease in black African heterosexuals unaware of their \ac{HIV} infection was pronounced over 2012-2017, from 2,700 (95\% \ac{CrI} 2,200 to 3,400) to 1,200 (95\% \ac{CrI} 1,000 to 1,500, $p = 1$), corresponding to a halving in the prevalence of undiagnosed infection from 4.0 (95\% \ac{CrI} 3.3 to 4.9) to 1.7 (95\% \ac{CrI} 1.4 to 2.2) per 1,000 population (Table \ref{tab:postEst}). However, for the sub-group of clinic-attending black African heterosexuals aged 45-59 outside London, there was little evidence of a decrease in numbers undiagnosed (Figure \ref{fig:NundiagBAhets}, $p = 0.541$). In contrast, the size of this 45-59 sub-group increased over the same time period, from 3,200 (95\% \ac{CrI} 3,000 to 3,300) to 4,900 (95\% \ac{CrI} 4,800 to 5,000) attending a clinic for \ac{STI}-related needs. Taking this change in denominator into account, the decrease in prevalence of undiagnosed infection in this group was estimated to be from 46 (95\% \ac{CrI} 27 to 76) to 29 (95\% \ac{CrI} 17 to 47) per 1,000 population over 2012-2017.

\begin{sidewaysfigure}
\includegraphics[width = \textwidth]{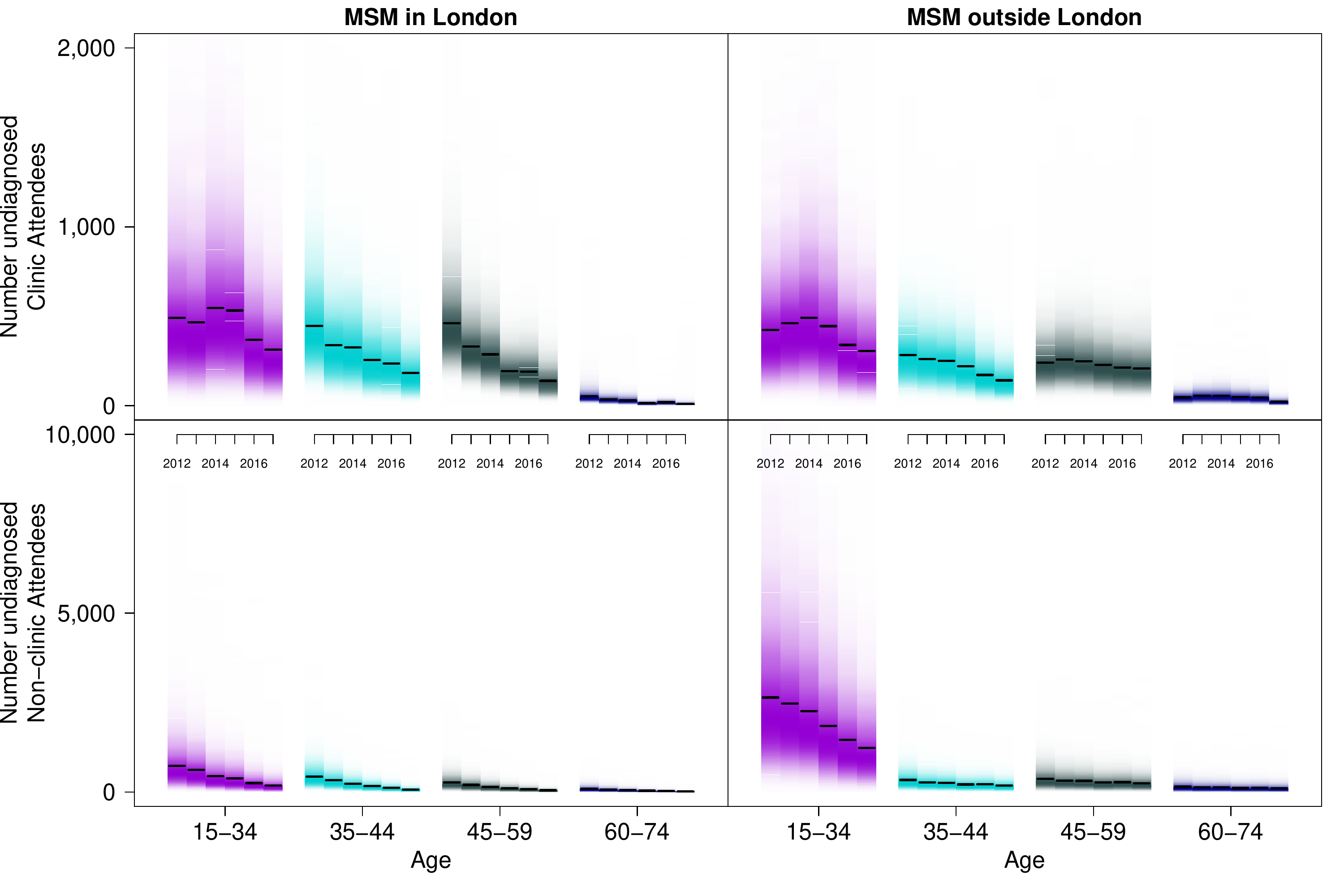}
\caption{Number of \ac{MSM} living with undiagnosed \ac{HIV}, by clinic attendance, region, age and year (2012-2017). Note the differing scales of the y-axes by clinic attendance.}
\label{fig:NundiagMSM}
\end{sidewaysfigure}

The decrease in numbers undiagnosed was notable in all sub-groups of \ac{MSM} (Figure \ref{fig:NundiagMSM}), with the total more than halving from 8,100 (95\% \ac{CrI} 4,700 to 14,800) in 2012 to 3,800 (95\% \ac{CrI} 2,100 to 7,300) in 2017 ($p = 0.965$, Table \ref{tab:postEst}). The corresponding prevalence of undiagnosed infection dropped from 16.0 (95\% \ac{CrI} 9.3 to 28.8) to 7.2 (95\% \ac{CrI} 3.9 to 13.8) per 1,000 population. Also, the decrease was more pronounced for all \ac{MSM}, in London ($p = 0.988$) compared to outside London ($p = 0.854$, Figure \ref{fig:NundiagMSM}). The numbers undiagnosed were largest, but also most uncertain, amongst the 15-34 age group, with posterior probabilities of a decrease lowest in the clinic-attending sub-group ($p = 0.714$ in London, $p = 0.662$ outside London). The disparity in trends between areas was greatest among 45-59 year olds, with posterior probabilities of a decrease estimated to be $p = 0.990$ in London, but only $p = 0.681$ outside London.

Due to small sample sizes, estimates of the number of \ac{PWID} living with undiagnosed \ac{HIV} are uncertain, with low posterior probability ($p = 0.26$) of any decrease over 2012-2017 (Table \ref{tab:postEst}). 
For all heterosexuals, trends in the number of \ac{PLWH} unaware of their infection were similar for men and women (Figures 6 and 7 in the Appendix). Overall, \ac{PLWH} were an ageing population in 2012-2017, with the prevalent burden concentrated in the 45-59 age group (Figure 8 of the Appendix). 

\section{Discussion \label{sec:Discussion}}

\ac{MSM}, \ac{PWID} and black African heterosexuals remain disproportionately affected by undiagnosed \ac{HIV} infection per population 
in 2017 (Table \ref{tab:postEst}). However, we have demonstrated remarkable decreases from 2012 to 2017 in the prevalence of \ac{HIV} in these groups and the first 90 of the \ac{UNAIDS} 90-90-90 targets was met in 2016 among the population aged 15-74 in England. The number of \ac{PLWH} unaware of their infection halved to 6,900 (95\% \ac{CrI} 4,900 to 10,700) over the study period, with comparable reductions estimated in all \ac{MSM} and most black African heterosexual sub-groups. However, there are three less encouraging findings from our analysis. First, the number undiagnosed outside London is not decreasing as fast as in London. Second, trends in numbers undiagnosed were too uncertain in other ethnicity heterosexuals who were not recent sexual health clinic attendees to draw definitive conclusions. Finally, the numbers of recent clinic-attending other ethnicity heterosexuals living with undiagnosed \ac{HIV} are of comparable absolute magnitude to those who had not recently attended a clinic. This implies many opportunities for testing are being missed in clinic attendees. Indeed, the latest \ac{PHE} report on \ac{HIV} testing \cite{PHEtesting2017} found that of eligible clinic attendees who were not either \ac{MSM}, black African or born in a high prevalence country, the proportion who were declining a \ac{HIV} test had increased over the previous five years, to 27\% in 2016. 

This analysis used the most recent available datasets to provide estimates of the latest trends in the \ac{HIV} epidemic in England. A key strength of our \ac{MPES} approach is the continued and improved ability to estimate the unobserved burden of \ac{HIV}, and in particular, to quantify temporal changes, which are critical to prioritising policies and monitoring progress towards \ac{HIV} elimination \cite{BHIVA2008,NICE2017,NAT2013,HPEcampaigns2019}. 
Our estimates rely on model assumptions necessary to identify unobservable quantities, including: relating undiagnosed prevalence to the proportion of \ac{HIV} tests giving new diagnoses; and smoothing constraints to address data sparsity. These assumptions are judged plausible, particularly as robustness is ensured by appropriately allowing for uncertainty: an example is the modelling of the dynamic nature of prevalence estimated from the sexual health clinic data within each year, assumed to lie between start- and end-year prevalences.

A consequence of both the epidemic
and available data sources evolving over time is
the continuing
adaptation
of the \ac{MPES} model. One outstanding issue is that the opiate-using population, including \ac{PWID}, is thought to be an ageing population \cite{ACMD2019}, so that the age-gender distribution assumed may be outdated. However, given the low and uncertain estimates of absolute numbers of \ac{PLWH} among \ac{PWID}, our estimates are relatively robust to this ageing. Changes in migration and other population patterns may also have occurred such that group sizes have changed since the \ac{ONS} Census and the \ac{NATSAL} survey were carried out in 2011. Newer data sources are therefore being sought to supplement the evidence base in more recent years, with accompanying model development to make better, more efficient use of existing and new data sources. This ongoing work includes: a new round of the \ac{NATSAL} survey in 2020/21 
(\url{www.natsal.ac.uk/online-consultation/background-to-natsal-4.aspx}); updating estimates of the \ac{PWID} population size; incorportion of information from community and online surveys \cite{Weatherburn2013,Logan2019}; and extending the \ac{MPES} model to datasets collected by the other \ac{UK} nations. 

Other 
approaches have been used to estimate trends in the number of undiagnosed \ac{HIV} infections in the \ac{UK}. First, the CD4-staged ``back-calculation'' approach \cite{Birrell2013a,Brizzi2019}, 
also a Bayesian evidence synthesis, based on different data sources, is used annually to estimate and monitor \ac{HIV} incidence in \ac{MSM} \cite{PHEHIVreport2018}. The corresponding estimates of the number undiagnosed are consistent with our \ac{MPES} estimates 
and 
both approaches provide complementary pictures of the \ac{HIV} epidemic in \ac{MSM} in England.
Another approach is transmission modelling \cite{Punyacharoensin2016,Cambiano2018}, aimed at forecasting both the epidemic and the effects of different possible interventions, particularly in the \ac{MSM} population
. Estimates of the number of \ac{PLWH} resulting from these transmission models are broadly consistent with our estimates. The \ac{UNAIDS} EPP/Spectrum software for estimating \ac{HIV} prevalence, incidence and mortality curves \cite{UNAIDS_SpectrumEPP2018}
, are also based on modelling transmission and demographic dynamics, but in contrast to the \ac{MPES} approach, do not provide 
estimates of undiagnosed \ac{HIV} prevalence. 

Our estimates have important implications for efforts to eliminate \ac{HIV} transmission in England, especially the challenge of diagnosing \ac{HIV} infections in other ethnicity heterosexuals. Compared to black African heterosexuals, there are now probably more undiagnosed HIV infections in other ethnicity heterosexuals, especially so outside of London. Moreover, the rate of decrease of undiagnosed HIV infections in other ethnicity heterosexuals who are not recent sexual health clinic attendees is the slowest of all. This might be expected, given members of this population sub-set may have no particular reason to consider themselves at \ac{HIV} risk -- if they did, then they could be expected to attend a clinic for a test. On the other hand, since undiagnosed \ac{HIV} prevalence in other ethnicity heterosexuals who are recent clinic attendees is almost 40 times greater than in those who are not (Table \ref{tab:postEstNBAbyAge}), the priority must be to ensure all of these have an \ac{HIV} test, rather than the 27\% that currently do so \cite{PHEtesting2017}. As all clinic attendees living with \ac{HIV} become diagnosed, improved partner notification \cite{Rayment2017BMJSTI} is likely to accelerate reducing the undiagnosed fraction in the wider population, making the prospect of \ac{HIV} elimination increasingly likely.

\section*{Acknowledgements}
The authors thank Dr Stefano Conti (NHS England), Louise Logan, Dr Stephanie Migchelsen, Katy Davison, Sarika Desai, Cuong Chau, Zheng Yin, Martina Furegato, Sophie Nash, Dr Nicky Connor (PHE), Prof. Cath Mercer (UCL), Dr Ford Hickson (LSHTM, Sigma Research) and Prof. Claire Thorne (UCL ICH) for providing data and constructive discussion. Acknowledgements are also due to all participants of a stakeholder engagement event, whose contributions to the discussions informed model developments aimed at making greater and more efficient use of available data. The event, held at London School of Hygiene and Tropical Medicine in 2016, was co-ordinated by Catherine Dodds (LSHTM and Sigma Research), Deborah Gold (National AIDS Trust), Valerie Delpech (PHE) and Daniela De Angelis (MRC Biostatistics Unit). 

\section*{Contributors}
AMP, DDA, ONG and VD conceived and designed the study. PK, SC, AM, HM, EH and AB collated the data and undertook exploratory data analysis. AMP independently verified all collated data, undertook the primary analysis, and drafted the manuscript. AMP, RH and CJ contributed to model development and coding, in discussion and consultation with all other authors. All authors contributed to subsequent iterations of the manuscript, provided critical input on the manuscript and approved the final version for publication. AMP is guarantor. The corresponding author attests that all listed authors meet authorship criteria and that no others meeting the criteria have been omitted.

\section*{Funding}
AMP, CJ and DDA were funded by the UK Medical Research Council programme MRC\_MC\_UU\_00002/11. DDA was additionally funded by Public Health England. PK, AM, SC, RH, EH, HM, AB, VD and ONG were employed by Public Health England. Funders did not have any role in the study design; the collection, analysis, and interpretation of data; the writing of the report; or the decision to submit the article for publication. AMP and DDA had access to the Public Health England data under honorary contracts with Public Health England, and all authors had full access to all of the data in the study and can take responsibility for the integrity of the data and the accuracy of the data analysis.

\section*{Data sharing}
Relevant data on which this analysis is based are available on request to Public Health England in accordance with Public Health England's HIV/STI Data Sharing Policy at \url{https://www.gov.uk/government/publications/hiv-and-sti-data-sharing-policy}. All requests for data access will need to specify the planned use of data and will require approval from Public Health England before release.

\begin{sidewaystable}
\resizebox{\textwidth}{!}{%
\begin{tabular}{ccrrP{1.35cm}rrP{1.35cm}rrrP{1.15cm}rrP{1.15cm}}
\toprule
	&		&	\multicolumn{7}{c}{\textbf{Number unaware of their HIV infection}}																									&	\multicolumn{6}{c}{\textbf{Undiagnosed prevalence per 100,000 population}}																							\\
	&		&	\multicolumn{3}{c}{\textbf{2012}}											&	\multicolumn{3}{c}{\textbf{2017}}											&		&	\multicolumn{3}{c}{\textbf{2012}}											&	\multicolumn{3}{c}{\textbf{2017}}											\\
\parbox{2.7cm}{\centering \ea{Exposure group}}	&	\ea{Region}	&	\ea{Median}	&	\multicolumn{2}{c}{\ea{95\% CrI}}									&	\ea{Median}	&	\multicolumn{2}{c}{\ea{95\% CrI}}									&	\ea{$\bs{p}$}	&	\ea{Median}	&	\multicolumn{2}{c}{\ea{95\% CrI}}									&	\ea{Median}	&	\multicolumn{2}{c}{\ea{95\% CrI}}									\\
\midrule																																									
\multirow{3}{*}{\parbox{2.7cm}{\centering MSM}}	&	London	&	3,226	&	(	1,765	-	&	6,216	)	&	1,062	&	(	532	-	&	2,238	)	&	0.99	&	2,253	&	(	1,239	-	&	4,304	)	&	695	&	(	352	-	&	1,472	)	\\
	&	Outside London	&	4,726	&	(	2,192	-	&	10,683	)	&	2,649	&	(	1,190	-	&	5,909	)	&	0.85	&	1,301	&	(	607	-	&	2,874	)	&	706	&	(	321	-	&	1,564	)	\\
	&	England	&	8,140	&	(	4,741	-	&	14,812	)	&	3,784	&	(	2,075	-	&	7,317	)	&	0.96	&	1,605	&	(	931	-	&	2,880	)	&	717	&	(	394	-	&	1,376	)	\\
\midrule																																									
\multirow{3}{*}{\parbox{2.7cm}{\centering PWID}}	&	London	&	50	&	(	17	-	&	121	)	&	70	&	(	21	-	&	176	)	&	0.32	&	353	&	(	121	-	&	851	)	&	476	&	(	145	-	&	1,201	)	\\
	&	Outside London	&	66	&	(	23	-	&	155	)	&	107	&	(	34	-	&	266	)	&	0.24	&	66	&	(	23	-	&	157	)	&	106	&	(	33	-	&	269	)	\\
	&	England	&	117	&	(	42	-	&	272	)	&	179	&	(	58	-	&	425	)	&	0.26	&	103	&	(	37	-	&	239	)	&	155	&	(	50	-	&	373	)	\\
\midrule																																									
\multirow{3}{*}{\parbox{2.7cm}{\centering heterosexuals (black African)}}	&	London	&	1,305	&	(	1,035	-	&	1,665	)	&	460	&	(	348	-	&	623	)	&	1.00	&	327	&	(	259	-	&	417	)	&	110	&	(	83	-	&	149	)	\\
	&	Outside London	&	1,426	&	(	1,096	-	&	1,857	)	&	735	&	(	566	-	&	964	)	&	1.00	&	503	&	(	387	-	&	654	)	&	258	&	(	199	-	&	339	)	\\
	&	England	&	2,738	&	(	2,232	-	&	3,363	)	&	1,198	&	(	962	-	&	1,520	)	&	1.00	&	401	&	(	327	-	&	493	)	&	171	&	(	137	-	&	217	)	\\
\midrule																																									
\multirow{3}{*}{\parbox{2.7cm}{\centering heterosexuals (other ethnicities)}}	&	London	&	841	&	(	595	-	&	1,475	)	&	399	&	(	263	-	&	832	)	&	0.97	&	15	&	(	10	-	&	26	)	&	7	&	(	4	-	&	14	)	\\
	&	Outside London	&	1,413	&	(	991	-	&	2,681	)	&	1,170	&	(	768	-	&	2,517	)	&	0.71	&	4	&	(	3	-	&	8	)	&	3	&	(	2	-	&	7	)	\\
	&	England	&	2,257	&	(	1,680	-	&	4,010	)	&	1,572	&	(	1,081	-	&	3,247	)	&	0.86	&	6	&	(	4	-	&	10	)	&	4	&	(	3	-	&	8	)	\\
\midrule																																																																	
\multirow{3}{*}{\parbox{2.7cm}{\centering all heterosexuals}}	&	London	&	2,170	&	(	1,766	-	&	2,869	)	&	871	&	(	673	-	&	1,324	)	&	1.00	&	35	&	(	29	-	&	47	)	&	13	&	(	10	-	&	20	)	\\
	&	Outside London	&	2,866	&	(	2,282	-	&	4,200	)	&	1,920	&	(	1,458	-	&	3,267	)	&	0.93	&	9	&	(	7	-	&	13	)	&	6	&	(	4	-	&	10	)	\\
	&	England	&	5,038	&	(	4,196	-	&	6,854	)	&	2,795	&	(	2,199	-	&	4,470	)	&	0.99	&	13	&	(	11	-	&	17	)	&	7	&	(	5	-	&	11	)	\\
\midrule																																									
\multirow{3}{*}{\parbox{2.7cm}{\centering Total}}	&	London	&	5,500	&	(	3,922	-	&	8,525	)	&	2,046	&	(	1,434	-	&	3,297	)	&	1.00	&	87	&	(	62	-	&	135	)	&	31	&	(	22	-	&	50	)	\\
	&	Outside London	&	7,760	&	(	5,062	-	&	13,706	)	&	4,791	&	(	3,128	-	&	8,213	)	&	0.92	&	23	&	(	15	-	&	41	)	&	14	&	(	9	-	&	24	)	\\
	&	England	&	13,452	&	(	9,813	-	&	20,184	)	&	6,913	&	(	4,921	-	&	10,727	)	&	0.99	&	34	&	(	25	-	&	51	)	&	17	&	(	12	-	&	26	)	\\
	
\bottomrule
\end{tabular}}
\caption{Posterior estimates by exposure group and region of the number and prevalence per 100,000 population of undiagnosed infections, in 2012 compared to 2017: posterior median, 95\% \ac{CrI} and posterior probability $p$ of a decrease over the five-year period. \label{tab:postEst}}
\end{sidewaystable}

\begin{sidewaystable}
\resizebox{\textwidth}{!}{%
\begin{tabular}{cccrrP{1.35cm}rrP{1.35cm}rrrP{1.15cm}rrP{1.15cm}}
\toprule
	&		&		&	\multicolumn{7}{c}{\textbf{Number unaware of their HIV infection}}																			&	\multicolumn{6}{c}{\textbf{Undiagnosed prevalence per 100,000 population}}																	\\
	&		&		&	\multicolumn{3}{c}{\textbf{2012}}								&	\multicolumn{3}{c}{\textbf{2017}}								&		&	\multicolumn{3}{c}{\textbf{2012}}								&	\multicolumn{3}{c}{\textbf{2017}}								\\
\parbox{3.7cm}{\centering \ea{Exposure group}}	&	\ea{Age}	&	\ea{Region}	&	\ea{Median}	&	\multicolumn{2}{c}{\ea{95\% CrI}}						&	\ea{Median}	&	\multicolumn{2}{c}{\ea{95\% CrI}}						&	$\bs{p}$	&	\ea{Median}	&	\multicolumn{2}{c}{\ea{95\% CrI}}						&	\ea{Median}	&	\multicolumn{2}{c}{\ea{95\% CrI}}						\\
\midrule																																											
\multirow{15}{*}{\parbox{3.7cm}{\centering Recent clinic-attending heterosexuals (other ethnicities)}}
	&	15-34	&	London	&	187	&	(	113	-	&	308	)	&	95	&	(	56	-	&	156	)	&	0.97	&	80	&	(	48	-	&	132	)	&	40	&	(	24	-	&	66	)	\\
	&		&	Outside London	&	333	&	(	210	-	&	523	)	&	244	&	(	162	-	&	364	)	&	0.85	&	57	&	(	36	-	&	90	)	&	34	&	(	23	-	&	51	)	\\
	&		&	England	&	527	&	(	361	-	&	759	)	&	341	&	(	243	-	&	475	)	&	0.96	&	64	&	(	44	-	&	93	)	&	36	&	(	26	-	&	50	)	\\
																																											
	&	35-44	&	London	&	138	&	(	81	-	&	242	)	&	57	&	(	31	-	&	108	)	&	0.98	&	319	&	(	187	-	&	558	)	&	117	&	(	64	-	&	223	)	\\
	&		&	Outside London	&	233	&	(	146	-	&	376	)	&	184	&	(	116	-	&	292	)	&	0.76	&	247	&	(	155	-	&	398	)	&	144	&	(	91	-	&	229	)	\\
	&		&	England	&	378	&	(	262	-	&	548	)	&	244	&	(	167	-	&	360	)	&	0.95	&	275	&	(	190	-	&	398	)	&	139	&	(	95	-	&	205	)	\\
																																											
	&	45-59	&	London	&	157	&	(	94	-	&	283	)	&	68	&	(	39	-	&	116	)	&	0.99	&	745	&	(	445	-	&	1,340	)	&	320	&	(	186	-	&	544	)	\\
	&		&	Outside London	&	234	&	(	140	-	&	394	)	&	169	&	(	101	-	&	282	)	&	0.81	&	388	&	(	232	-	&	653	)	&	212	&	(	127	-	&	352	)	\\
	&		&	England	&	397	&	(	274	-	&	591	)	&	240	&	(	163	-	&	360	)	&	0.96	&	489	&	(	337	-	&	727	)	&	237	&	(	161	-	&	355	)	\\
																																											
	&	60-74	&	London	&	54	&	(	32	-	&	91	)	&	17	&	(	8	-	&	32	)	&	1.00	&	1,582	&	(	938	-	&	2,685	)	&	461	&	(	229	-	&	872	)	\\
	&		&	Outside London	&	50	&	(	28	-	&	86	)	&	54	&	(	31	-	&	93	)	&	0.42	&	452	&	(	253	-	&	783	)	&	375	&	(	216	-	&	647	)	\\
	&		&	England	&	106	&	(	71	-	&	155	)	&	72	&	(	46	-	&	112	)	&	0.90	&	732	&	(	496	-	&	1,073	)	&	399	&	(	257	-	&	623	)	\\
\cmidrule{2-16}
	&	Total	&	London	&	550	&	(	415	-	&	740	)	&	243	&	(	180	-	&	331	)	&	1.00	&	182	&	(	138	-	&	245	)	&	79	&	(	58	-	&	107	)	\\
	&		&	Outside London	&	868	&	(	663	-	&	1,144	)	&	663	&	(	520	-	&	850	)	&	0.93	&	116	&	(	88	-	&	153	)	&	71	&	(	56	-	&	91	)	\\
	&		&	England	&	1,424	&	(	1,159	-	&	1,763	)	&	909	&	(	746	-	&	1,112	)	&	1.00	&	135	&	(	110	-	&	168	)	&	73	&	(	60	-	&	89	)	\\
\midrule																																											
\multirow{15}{*}{\parbox{3.7cm}{\centering Non-clinic-attending heterosexuals (other ethnicities)}}
	&	15-34	&	London	&	129	&	(	33	-	&	312	)	&	70	&	(	15	-	&	185	)	&	0.78	&	6	&	(	1	-	&	14	)	&	3	&	(	1	-	&	8	)	\\
	&		&	Outside London	&	272	&	(	73	-	&	656	)	&	198	&	(	43	-	&	509	)	&	0.67	&	3	&	(	1	-	&	6	)	&	2	&	(	0	-	&	5	)	\\
	&		&	England	&	410	&	(	121	-	&	900	)	&	275	&	(	63	-	&	647	)	&	0.72	&	3	&	(	1	-	&	7	)	&	2	&	(	1	-	&	5	)	\\
																																											
	&	35-44	&	London	&	75	&	(	4	-	&	411	)	&	38	&	(	3	-	&	266	)	&	0.67	&	7	&	(	0	-	&	37	)	&	3	&	(	0	-	&	22	)	\\
	&		&	Outside London	&	110	&	(	6	-	&	634	)	&	122	&	(	8	-	&	684	)	&	0.47	&	2	&	(	0	-	&	11	)	&	2	&	(	0	-	&	13	)	\\
	&		&	England	&	190	&	(	12	-	&	1,005	)	&	165	&	(	12	-	&	919	)	&	0.54	&	3	&	(	0	-	&	15	)	&	2	&	(	0	-	&	14	)	\\
																																											
	&	45-59	&	London	&	44	&	(	3	-	&	249	)	&	23	&	(	2	-	&	164	)	&	0.66	&	3	&	(	0	-	&	19	)	&	2	&	(	0	-	&	11	)	\\
	&		&	Outside London	&	85	&	(	5	-	&	529	)	&	106	&	(	7	-	&	628	)	&	0.44	&	1	&	(	0	-	&	6	)	&	1	&	(	0	-	&	7	)	\\
	&		&	England	&	135	&	(	9	-	&	736	)	&	132	&	(	9	-	&	760	)	&	0.50	&	1	&	(	0	-	&	7	)	&	1	&	(	0	-	&	7	)	\\
																																											
	&	60-74	&	London	&	13	&	(	1	-	&	79	)	&	7	&	(	0	-	&	50	)	&	0.67	&	2	&	(	0	-	&	10	)	&	1	&	(	0	-	&	6	)	\\
	&		&	Outside London	&	34	&	(	2	-	&	211	)	&	42	&	(	3	-	&	262	)	&	0.44	&	0	&	(	0	-	&	3	)	&	1	&	(	0	-	&	4	)	\\
	&		&	England	&	50	&	(	3	-	&	274	)	&	51	&	(	4	-	&	301	)	&	0.49	&	1	&	(	0	-	&	4	)	&	1	&	(	0	-	&	4	)	\\
\cmidrule{2-16}																																											
	&	Total	&	London	&	277	&	(	82	-	&	900	)	&	150	&	(	41	-	&	578	)	&	0.76	&	5	&	(	2	-	&	17	)	&	3	&	(	1	-	&	10	)	\\
	&		&	Outside London	&	531	&	(	164	-	&	1,775	)	&	498	&	(	136	-	&	1,827	)	&	0.54	&	2	&	(	1	-	&	6	)	&	2	&	(	0	-	&	6	)	\\
	&		&	England	&	817	&	(	264	-	&	2,571	)	&	655	&	(	189	-	&	2,313	)	&	0.61	&	2	&	(	1	-	&	7	)	&	2	&	(	0	-	&	6	)	\\
\bottomrule
\end{tabular}}
\caption{Posterior estimates by age group, clinic attendance and region of the number and prevalence per 100,000 population of undiagnosed infections in other ethnicty heterosexuals, in 2012 compared to 2017: posterior median, 95\% \ac{CrI} and posterior probability $p$ of a decrease over the five-year period. \label{tab:postEstNBAbyAge}}
\end{sidewaystable}

\newpage
\clearpage

\bibliographystyle{vancouver}
\bibliography{mpesRefs}

\begin{thebibliography}{10}

\bibitem{Cohen2011}
Cohen MS, Chen YQ, McCauley M, Gamble T, Hosseinipour MC, Kumarasamy N, et~al.
\newblock {Prevention of HIV-1 infection with early antiretroviral therapy}.
\newblock New England Journal of Medicine. 2011;365(6):493--505.

\bibitem{GranichEtAl2009}
Granich RM, Gilks CF, Dye C, {De Cock} KM, Williams BG.
\newblock {Universal voluntary HIV testing with immediate antiretroviral
  therapy as a strategy for elimination of HIV transmission: a mathematical
  model}.
\newblock The Lancet. 2009;373(9657):48--57.
\newblock Available from:
  \url{http://dx.doi.org/10.1016/s0140-6736(08)61697-9}.

\bibitem{UNAIDS9090902014}
{UN Joint Programme on HIV/AIDS (UNAIDS)}.
\newblock {90-90-90 An ambitious treatment target to help end the AIDS
  epidemic}; 2014.
\newblock Available from:
  \url{https://www.unaids.org/sites/default/files/media{\_}asset/90-90-90{\_}en.pdf,
  accessed 23rd September 2019}.

\bibitem{Churchill2016}
Churchill D, Waters L, Ahmed N, Angus B, Boffito M, Bower M, et~al.
\newblock {British HIV Association guidelines for the treatment of
  HIV-1-positive adults with antiretroviral therapy 2015}.
\newblock HIV Medicine. 2016;17:s2--s104.

\bibitem{McCormack2016}
McCormack S, Dunn DT, Desai M, Dolling DI, Gafos M, Gilson R, et~al.
\newblock {Pre-exposure prophylaxis to prevent the acquisition of HIV-1
  infection (PROUD): effectiveness results from the pilot phase of a pragmatic
  open-label randomised trial}.
\newblock The Lancet. 2016;387:53--60.
\newblock Available from:
  \url{http://dx.doi.org/10.1016/S0140-6736(15)00056-2}.

\bibitem{PHEHIVreport2018}
Nash S, Desai S, Croxford S, Guerra L, Lowndes C, Connor N, et~al.
\newblock {Progress towards ending the HIV epidemic in the United Kingdom 2018
  report}.
\newblock Public Health England; 2018. December 2017.

\bibitem{ONSCensus2012}
{Office for National Statistics}.
\newblock {2011 Census: population and household estimates for England and
  Wales}.
\newblock Office for National Statistics; 2012.
\newblock Accessed 29th April 2019.
\newblock Available from:
  \url{https://www.ons.gov.uk/peoplepopulationandcommunity/populationandmigration/populationestimates/bulletins/2011censuspopulationandhouseholdestimatesforenglandandwales/2012-07-16/pdf}.

\bibitem{BHIVA2008}
{British HIV Association}, {British Association of Sexual Health \& HIV},
  {British Infection Society}.
\newblock {UK National Guidelines for HIV Testing 2008}; 2008. September.
\newblock Available from:
  \url{https://www.bhiva.org/file/RHNUJgIseDaML/GlinesHIVTest08.pdf, accessed
  23rd September 2019.}

\bibitem{NICE2017}
{National Institute for Health and Care Excellence}.
\newblock {HIV testing: encouraging uptake}; 2017. September.
\newblock Available from: \url{www.nice.org.uk/guidance/qs157, accessed 23rd
  September 2019.}

\bibitem{NAT2013}
{National AIDS Trust}.
\newblock {A practical guide for Commissioners Commissioning HIV Testing
  Services In England}; 2013. November.
\newblock Available from:
  \url{https://www.nat.org.uk/system/files{\_}force/publications/Nov{\_}2013{\_}Toolkit.pdf,
  accessed 23rd September 2019.}

\bibitem{HPEcampaigns2019}
{HIV Prevention England}.
\newblock {``It Starts With Me'' and ``National HIV Testing Week''}.
\newblock {HIV Prevention England}; 2019.
\newblock Available from:
  \url{https://www.hivpreventionengland.org.uk/campaigns/, accessed 23rd
  September 2019.}

\bibitem{Goubar2008}
Goubar A, Ades AE, {De Angelis} D, McGarrigle CA, Mercer CH, Tookey PA, et~al.
\newblock {Estimates of human immunodeficiency virus prevalence and proportion
  diagnosed based on Bayesian multiparameter synthesis of surveillance data}.
\newblock Journal of the Royal Statistical Society Series A: Statistics in
  Society. 2008;171(3):541--580.

\bibitem{DeAngelis2014}
{De Angelis} D, Presanis AM, Conti S, Ades AE.
\newblock {Estimation of HIV burden through Bayesian evidence synthesis}.
\newblock Statistical Science. 2014;29(1):9--17.

\bibitem{rStan2018}
{Stan Development Team}. {RStan}: the {R} interface to {Stan}; 2018.
\newblock R package version 2.17.3.
\newblock Available from: \url{http://mc-stan.org/}.

\bibitem{ONS_MYEs2018}
{Office for National Statistics}.
\newblock {Population Estimates for UK, England and Wales Scotland, and
  Northern Ireland Mid-2017 Population Estimates.}; 2018. July.
\newblock Accessed 29th April 2019.
\newblock Available from: \url{www.nomisweb.co.uk
  https://www.ons.gov.uk/peoplepopulationandcommunity/populationandmigration/populationestimates/bulletins/annualmidyearpopulationestimates/mid2017/pdf}.

\bibitem{Mercer2016}
Mercer CH, Prah P, Field N, Tanton C, Macdowall W, Clifton S, et~al.
\newblock {The health and well-being of men who have sex with men ( MSM ) in
  Britain : Evidence from the third National Survey of Sexual Attitudes and
  Lifestyles ( Natsal-3 )}.
\newblock BMC Public Health. 2016;p. 1--16.

\bibitem{HayEtAl2011}
Hay G, Gannon M, Casey J, Millar T.
\newblock {Estimates of the Prevalence of Opiate Use and/or Crack Cocaine Use,
  2009/10: Sweep 6 report}.
\newblock Centre for Public Health, Liverpool John Moores University; 2011. 6.
\newblock Accessed 29th April 2019.
\newblock Available from:
  \url{http://citeseerx.ist.psu.edu/viewdoc/download?doi=10.1.1.690.2170{\&}rep=rep1{\&}type=pdf}.

\bibitem{KingEtAl2014}
King R, Bird SM, Hay G, Hutchingson SJ.
\newblock {Estimating Prevalence of Injecting Drug Users and Associated Death
  Rates in England Using Regional Data and Incorporating Prior Information}.
\newblock Journal of the Royal Statistical Society: Series A (Statistics in
  Society). 2014;177(1):209--236.
\newblock Available from:
  \url{https://rss.onlinelibrary.wiley.com/doi/pdf/10.1111/rssa.12011}.

\bibitem{SweetingEtAl2009}
Sweeting MJ, {De Angelis} D, Ades AE, Hickman M.
\newblock {Estimating the prevalence of ex-injecting drug use in the
  population}.
\newblock Statistical Methods in Medical Research. 2009;18(4):381--395.
\newblock Available from: \url{http://dx.doi.org/10.1177/0962280208094704}.

\bibitem{Savage2014}
Savage EJ, Mohammed H, Leong G, Duffell S, Hughes G.
\newblock {Improving surveillance of sexually transmitted infections using
  mandatory electronic clinical reporting: The genitourinary medicine clinic
  activity dataset, England, 2009 to 2013}.
\newblock Eurosurveillance. 2014;19(48):1--9.

\bibitem{Aghaizu2016}
Aghaizu A, Wayal S, Nardone A, Parson V, Copas AJ, Mercey DE, et~al.
\newblock {Understanding continuing high HIV incidence: trends in sexual
  behaviours, HIV testing and the proportion of men at risk of transmitting and
  acquiring HIV in London 2000-2013. A serial cross-sectional study.}
\newblock The Lancet HIV. 2016;3(9):e431--e440.

\bibitem{PublicHealthEngland2018}
{Public Health England}.
\newblock {Shooting Up: Infections among people who inject drugs in the UK,
  2017}.
\newblock Public Health England; 2018.
\newblock Accessed 29th April 2019.
\newblock Available from:
  \url{https://assets.publishing.service.gov.uk/government/uploads/system/uploads/attachment\_data/file/756502/Shooting\_up\_2018.pdf}.

\bibitem{Peters2018}
Peters H, Thorne C, Tookey PA, Byrne L.
\newblock {National audit of perinatal HIV infections in the UK, 2006–2013:
  what lessons can be learnt?}
\newblock HIV Medicine. 2018;19(4):280--289.

\bibitem{ONSliveBirthsRoB2018}
{Office for National Statistics}.
\newblock {Births in England and Wales by Parents' Country of Birth}.
\newblock Office for National Statistics; 2018.
\newblock Accessed 29th April 2019.
\newblock Available from:
  \url{https://www.ons.gov.uk/peoplepopulationandcommunity/birthsdeathsandmarriages/livebirths/bulletins/parentscountryofbirthenglandandwales/2017/pdf}.

\bibitem{Bourne2014}
Bourne A, Reid D, Weatherburn P.
\newblock {African Health and Sex Survey 2013-2014: headline findings}.
\newblock London, England: Sigma Research, London School of Hygiene and
  Tropical Medicine; 2014.
\newblock Accessed 29th April 2019.
\newblock Available from:
  \url{http://www.sigmaresearch.org.uk/files/report2014c.pdf}.

\bibitem{PHE_NHSBT_2017}
{National Health Service Blood and Transplant}, {Public Health England}.
\newblock {Safe supplies 2017: data sources and methods}.
\newblock NHS Blood and Transplant, Public Health England; 2017.
\newblock Accessed 29th April 2019.
\newblock Available from:
  \url{https://www.gov.uk/government/publications/safe-supplies-annual-review}.

\bibitem{Rice2017}
Rice BD, Yin Z, Brown AE, Croxford S, Conti S, {De Angelis} D, et~al.
\newblock {Monitoring of the HIV Epidemic Using Routinely Collected Data: The
  Case of the United Kingdom}.
\newblock AIDS and Behavior. 2017;21(s1):83--90.

\bibitem{PHEtesting2017}
Nash S, Furegato M, Gill O, Connor N, {contributors}.
\newblock {HIV Testing in England: 2017}; 2017.
\newblock Available from:
  \url{https://assets.publishing.service.gov.uk/government/uploads/system/uploads/attachment\_data/file/759270/HIV\_testing\_in\_England\_2017\_report.pdf}.

\bibitem{ACMD2019}
{Advisory Council on the Misuse of Drugs}.
\newblock {ACMD Report 2019: Ageing cohort of drug users}.
\newblock Advisory Council on the Misuse of Drugs; 2019.
\newblock Available from:
  \url{https://www.gov.uk/government/uploads/system/uploads/attachment{\_}data/file/237037/ACMD{\_}advice{\_}Z{\_}drugs.pdf}.

\bibitem{Weatherburn2013}
Weatherburn P, Schmidt AJ, Hickson F, Reid D, Berg RC, Hospers HJ, et~al.
\newblock {The European men-who-have-sex-with-men internet survey (EMIS):
  Design and methods}.
\newblock Sexuality Research and Social Policy. 2013;10(4):243--257.

\bibitem{Logan2019}
Logan L, Fakoya I, Howarth A, Murphy G, Johnson AM, Rodger AJ, et~al.
\newblock {Combination prevention and HIV: a cross-sectional community survey
  of gay and bisexual men in London, October to December 2016}.
\newblock Euro surveillance : bulletin Europeen sur les maladies transmissibles
  = European communicable disease bulletin. 2019;24(25).

\bibitem{Birrell2013a}
Birrell PJ, Gill ON, Delpech VC, Brown AE, Desai S, Chadborn TR, et~al.
\newblock {HIV incidence in men who have sex with men in England and Wales
  2001-10: A nationwide population study}.
\newblock The Lancet Infectious Diseases. 2013;13(4):313--318.
\newblock Available from:
  \url{http://dx.doi.org/10.1016/S1473-3099(12)70341-9}.

\bibitem{Brizzi2019}
Brizzi F, Birrell PJ, Plummer MT, Kirwan P, Brown AE, Delpech VC, et~al.
\newblock {Extending Bayesian back-calculation to estimate age and time
  specific HIV incidence}.
\newblock Lifetime Data Analysis. 2019;Available from:
  \url{https://doi.org/10.1007/s10985-019-09465-1}.

\bibitem{Punyacharoensin2016}
Punyacharoensin N, Edmunds W, {De Angelis} D, Delpech V, Hart G, Elford J,
  et~al.
\newblock {Effect of pre-exposure prophylaxis and combination HIV prevention
  for men who have sex with men in the UK: A mathematical modelling study}.
\newblock The Lancet HIV. 2016;3(2):e94--e104.

\bibitem{Cambiano2018}
Cambiano V, Miners A, Dunn D, McCormack S, Ong KJ, Gill ON, et~al.
\newblock {Cost-effectiveness of pre-exposure prophylaxis for HIV prevention in
  men who have sex with men in the UK: a modelling study and health economic
  evaluation}.
\newblock The Lancet Infectious Diseases. 2018;18(1):85--94.
\newblock Available from:
  \url{http://dx.doi.org/10.1016/S1473-3099(17)30540-6}.

\bibitem{UNAIDS_SpectrumEPP2018}
UNAIDS.
\newblock {Methods for deriving UNAIDS estimates 2018. Spectrum/EPP: Annex on
  Methods}; 2018.
\newblock Available from:
  \url{https://www.unaids.org/en/resources/documents/2018/Methods{\_}deriving{\_}estimates{\_}2018}.

\bibitem{Rayment2017BMJSTI}
Rayment M, Curtis H, Carne C, McClean H, Bell G, Estcourt C, et~al.
\newblock {An effective strategy to diagnose HIV infection: findings from a
  national audit of HIV partner notification outcomes in sexual health and
  infectious disease clinics in the UK}.
\newblock Sexually transmitted infections. 2017;93(2):94--99.
\newblock Available from: \url{https://doi.org/10.1136/sextrans-2015-052532}.

\end{thebibliography}

\section*{List of Acronyms}
\begin{acronym}
\acro{AHSS}{African Health and Sex Survey}
\acro{AIDS}{Acquired Immuno-Deficiency Syndrome}
\acro{ART}{anti-retroviral therapy}
\acro{BHIVA}{British HIV Association}
\acro{CrI}{credible interval}
\acro{EMIS}{European MSM Internet Survey}
\acro{GMSHS}{Gay Men's Sexual Health Survey}
\acro{GUMCAD}{Genito-Urinary Medicine Clinic Activity Dataset}
\acro{HARS}{HIV/AIDS Reporting System}
\acro{HIV}{Human Immuno-deficiency Virus}
\acro{HO}{Home Office}
\acro{MPES}{Multi-Parameter Evidence Synthesis}
\acro{MSM}{gay, bisexual and other men who have sex with men}
\acro{NATSAL}{National Survey of Sexual Attitudes and Lifestyles}
\acro{NHSBT}{National Health Service Blood and Transplant}
\acro{NSHPC}{National Study of HIV in Pregnancy and Childhood}
\acro{ONS}{Office for National Statistics}
\acro{PHE}{Public Health England}
\acro{PLWH}{people living with HIV}
\acro{PrEP}{\ac{HIV} pre-exposure prophylaxis}
\acro{PWID}{people who inject drugs}
\acro{SHC}{sexual health clinic}
\acro{SHS}{sexual health service}
\acro{SSA}{sub-Saharan Africa}
\acro{STI}{sexually transmitted infection}
\acro{TasP}{Treatment as Prevention}
\acro{UAM}{Unlinked Anonymous Monitoring}
\acro{UK}{United Kingdom}
\acro{UN}{United Nations}
\acro{UNAIDS}{Joint United Nations Programme on HIV/AIDS}
\acro{WHO}{World Health Organization}
\end{acronym}

\end{document}